\documentstyle[12pt]{article}
\newcommand{\thspace}{\kern.08333em}

\def \Bbar{\bar B}
\def \beqn{\begin{eqnarray}}
\def \beq{\begin{equation}}

\def \Kbar{\bar K}
\def \cn{Collaboration}

\def \eeq{\end{equation}}
\def \eeqn{\end{eqnarray}}

\def \s{\sqrt{2}}

\def \fbar{\bar{f}}

\def \v#1#2{V_{#1#2}}

\begin{document}

\rightline{CERN-TH/2000-250}
\rightline{August 2000}
\rightline{hep-ph/0008292}

\bigskip
\bigskip
\begin{center}
\Large\bf \boldmath U-spin Symmetry in Charmless $B$ Decays
\end{center}

\bigskip
\centerline{Michael Gronau\footnote{Permanent Address: Physics
Department,
Technion -- Israel Institute of Technology, 32000 Haifa, Israel.}}
\centerline{\it Theory Division, CERN}
\centerline{\it CH-1211, Geneva 23, Switzerland}

\vskip 1cm

\centerline{\bf ABSTRACT}
\bigskip

\begin{quote}
We prove a general theorem about equal CP rate differences within pairs
of U-spin related charmless $B$ and $B_s$ decays.
Large deviations from equalities would be evidence for new physics.
Six pairs of decays into
two pseudoscalar mesons are identified where such relations hold.
Ratios of corresponding rate differences and certain ratios of rates
measure U-spin breaking. These processes provide useful information on
the weak phase $\gamma={\rm Arg} V^*_{ub}$. Applications of U-spin
symmetry to other decays are discussed.

\end{quote}
\bigskip


Applying flavor symmetries of strong interactions to $B$ decays into two
light pseudoscalar mesons ($B\to PP$) may provide useful information
about phases of the
Cabibbo-Kobayashi-Maskawa (CKM) matrix. Using isospin symmetry, the three
$B\to\pi\pi$ decays, $B^0\to \pi^+\pi^-,~\pi^0\pi^0$ and $B^+\to\pi^+\pi^0$,
were shown \cite{GL} to determine quite precisely the weak phase
$\alpha={\rm Arg} (-V^*_{tb}V_{td}/V^*_{ub}V_{ud})$ (at least in principle,
albeit a potential difficulty in measuring $B^0\to\pi^0\pi^0$).
Within the approximation of flavor SU(3) symmetry, various aspects of
$B\to K\pi$ decays were studied \cite{BKpi} to learn the weak phase
$\gamma={\rm Arg}
(-V^*_{ub}V_{ud}/V^*_{cb}V_{cd})$. An accurate determination of
$\gamma$ requires the
knowledge of SU(3) breaking and rescattering effects which modify some of
the amplitudes.
Recently the important role of two $B_s$ decay processes,
$B_s\to K^+ K^-$ and $B_s\to K^-\pi^+$,  was demonstrated in this framework
\cite{FL, GR}. Here one makes explicit use only of a discrete U-spin symmetry
transformation interchanging $d$ and $s$ quarks \cite{U}.

The purpose of this Letter is to reconsider more generally the implications
of U-spin symmetry in charmless $B$ decays \cite{Lipkin}. First, we will
look at a very
general case of charmless $B^0,~B^+$ and $B_s$ decays, comparing decays from
and into any U-spin related states. We will prove as a general theorem
that {\it pairs of U-spin related processes involve CP rate differences which
are equal in magnitude and are opposite in sign}.
This property applies not only to two-body and quasi-two-body decays,
but in fact to any pair of U-spin related processes
including radiative decays.
Then, by focusing on $B\to PP$, we will find altogether twelve processes,
arranged in six pairs, where in each pair the decay amplitude of one process
is related to that of the other by U-spin symmetry, such that the CP rate
differences in the two processes are equal in magnitude. Several
processes lead to useful information on rescattering effects. A systematic
study of U-spin breaking, achieved by comparing some of these processes to
others, can lead to an accurate determination of $\gamma$.

The implications of U-spin symmetry in all charmless $B$ decays follow
from the following general considerations. The low energy effective weak
Hamiltonian describing $\Delta S =1$ $B$ decays is \cite{BBL}
\beq\label{Heff}
{\cal H}_{\rm eff} = \frac{G_F}{\s}\left[V^*_{ub}V_{us}\left(\sum^2_1 c_i
Q^{us}_i +\sum^{10}_3 c_i Q^s_i\right ) + V^*_{cb}V_{cs}\left(\sum^2_1 c_i
Q^{cs}_i +\sum^{10}_3 c_i Q^s_i\right )\right]~,
\eeq
where $c_i$ are scale-dependent Wilson coefficients and the flavor structure
of the various four-quark operators is $Q^{qs}_{1,2}\sim\bar b
q\bar q s,~Q^s_{3,..,6}\sim \bar b s \sum \bar q' q',~Q^s_{7,..,10}\sim
\bar b s\sum e_{q'}\bar q' q'~(q'=u,d,s,c$). Each of these operators
represents an $s$ component (``down") of a U-spin doublet, so that
one can write in short
\beq\label{Us}
{\cal H}_{\rm eff} = V^*_{ub}V_{us}U^s + V^*_{cb}V_{cs}C^s~,
\eeq
where $U$ and $C$ are U-spin doublet operators. Similarly, the effective
Hamiltonian responsible for $\Delta S =0$ decays involves $d$ components
(``up" in U-spin) of corresponding operators multiplying CKM factors
$V^*_{ub}V_{ud}$ and $V^*_{cb}V_{cd}$,
\beq\label{Ud}
{\cal H}_{\rm eff} = V^*_{ub}V_{ud}U^d + V^*_{cb}V_{cd}C^d~.
\eeq

Now, assume for simplicity \cite{other} that one compares two decay
processes, $\Delta S=1$ and
$\Delta S =0$, in which the initial and final states
are obtained from each other by a U-spin transformation,
$U: d\leftrightarrow s$.
Eqs.~(\ref{Us}) and (\ref{Ud}) imply that if the $\Delta S =1$ amplitude
(for the process $B \to f$) is written as
\beq\label{s}
A(B\to f,~\Delta S =1) = V^*_{ub}V_{us}A_u + V^*_{cb}V_{cs}A_c~,
\eeq
where $A_u$ and $A_c$ are complex amplitudes (involving CP-conserving phases),
then the corresponding $\Delta S =0$ amplitude (for $UB \to Uf$) is given by
\beq\label{d}
A(UB\to Uf,~\Delta S =0) = V^*_{ub}V_{ud}A_u + V^*_{cb}V_{cd}A_c~.
\eeq
The amplitudes of the  corresponding charge-conjugate processes are
\beq\label{sbar}
A(\Bbar \to \fbar,~\Delta S =-1) = V_{ub}V^*_{us}A_u + V_{cb}V^*_{cs}A_c~,
\eeq
and
\beq\label{dbar}
A(U\Bbar\to U\fbar,~\Delta S =0) = V_{ub}V^*_{ud}A_u + V_{cb}V^*_{cd}A_c~.
\eeq

To appreciate the powerful implication of U-spin symmetry in $B$
and $B_s$ decays, we note the following. The rates of the four processes
(\ref{s})$-$(\ref{dbar}) depend on four quantities, $\vert V^*_{ub}V_{us}
A_u\vert,~\vert V^*_{cb}V_{cs} A_c\vert,~\delta\equiv
{\rm Arg}(A_u A^*_c)$ and $\gamma\equiv {\rm Arg}(-V^*_{ub}V_{ud}V_{cb}
V^*_{cd})$.
Naively it may seem possible to use the corresponding four rates for a
determination of these four quantities including the weak phase $\gamma$.
However, there exists a general
U-spin relation between corresponding CP rate differences
\beq\label{asym}
\vert A(B \to f)\vert^2 - \vert A(\Bbar \to \fbar)\vert^2 =
-[\vert A(UB \to Uf)\vert ^2 - \vert A(U\Bbar \to U\fbar)\vert^2]~.
\eeq
This equality, which follows from the CKM unitarity relation \cite{Jarl}
\beq
{\rm Im}(V^*_{ub}V_{us}V_{cb}V^*_{cs}) =
- {\rm Im}(V^*_{ub}V_{ud}V_{cb}V^*_{cd})~,
\eeq
prohibits a determination of $\gamma$ from the four rates alone.
Towards the end of this Letter we will return to the question of determining
$\gamma$ by using another input measurement.

Eq.~(\ref{asym}), our general theorem for equal CP rate asymmetries, can be
used to test the validity and
accuracy of U-spin symmetry in {\it all types of $B$ decays}. These include
decays into two pseudoscalars (such as the pair $B^0\to K^+\pi^-$ and
$B_s\to\pi^+ K^-$), decays into a pseudoscalar and a vector meson
(e.g. $B^0\to K^{*+}\pi^-$ and $B_s\to\rho^+ K^-$), decays into two vector
mesons (e.g. $B^0 \to K^{*+}\rho^-$ and $B_s\to \rho^+ K^{*-}$) and
decays into multibody states (e.g. $B^+\to K^+ \pi^+\pi^-$ and
$B^+\to \pi^+ K^+ K^-$).
Our theorem applies also to radiative decays (e.g. $B^+\to K^{*+}
\gamma$ and $B^+\to \rho^+\gamma$) where the transition operators belong
to a U-spin doublet and involve suitable CKM factors \cite{BBL}.
A similar equality, between {\it inclusive} $b \to s\gamma$ and $b\to d\gamma$
CP rate differences was proved some time ago \cite{Soares}.
U-spin symmetry implies equal and opposite
sign CP rate differences within every pair of U-spin related processes.
Consequently, in each pair the process with the smaller rate is expected
to have a larger CP asymmetry. For instance,
the CP asymmetries in $B^+\to \Kbar^0 K^+$ (see discussion below)
and $B^+\to \rho^+\gamma$ are expected to be much larger than in $B^+\to
K^0\pi^+$ and $B^+\to K^{*+}\gamma$, respectively.

Of course, U-spin is only an approximate symmetry. Typically, U-spin breaking
is expected to be given by ratios of form factors and decay constants
(see discussion below). This does not lead to gross violation
of the above relations. {\it Large deviations from these equalities, and in
particular wrong relative signs of CP asymmetries \cite{sign}, would be
evidence for new physics.} Such physics would involve
additional flavor changing $b\to q~(q=d,s)$ transition operators which do not
have the U-spin and CKM structure of Eqs.~(\ref{Us}) and (\ref{Ud}).

In order to illustrate the consequences of this very general relation in
particular cases, we proceed to study in detail the overall implications
of U-spin symmetry in decays to two light pseudoscalar mesons.
Among all sixteen measurable $B$ meson decays \cite{GHLR} of the form
$B,B_s\to\pi\pi,~K\pi,~K\Kbar$ one can identify a dozen processes which
can be arranged in six U-spin related pairs:
\begin{enumerate}
\item $B^0 \to K^+\pi^-$~~vs.~~$B_s \to \pi^+ K^-$~,
\item $B_s \to K^+ K^-$~~vs.~~$B^0 \to \pi^+\pi^-$~,
\item $B^0 \to K^0\pi^0$~~vs.~~$B_s \to \Kbar^0\pi^0$~,
\item $B^+ \to K^0 \pi^+$~~vs.~~$B^+ \to \Kbar^0 K^+$~,
\item $B_s \to K^0 \Kbar^0$~~vs.~~$B^0 \to \Kbar^0 K^0$~,
\item $B_s \to\pi^+\pi^-$~~vs.~~$B^0 \to K^+ K^-$~.
\end{enumerate}
In case 3 the $\Delta U=1/2$ transition operators lead to $U=1$ final states
to which only the $U=0$ component of the $\pi^0$ contributes. (The two
pseudoscalar mesons are in an S-wave implying that they are in a symmetric
U-spin state).

Including $B_s \to \pi^0\pi^0$, which is related by isospin to $B_s\to \pi^+
\pi^-$, this consists of all but the following three $B\to PP$ ($P=\pi, K$)
decays: $B^+ \to K^+\pi^0,~B^+\to \pi^+\pi^0$ and $B^0\to\pi^0\pi^0$.
As our theorem Eq.~(\ref{asym}) states, in the U-spin symmetry limit the two
CP rate differences within each of the above six pairs of processes are equal
in magnitude and have opposite signs. Deviations from equal asymmetries
measure U-spin breaking.

For a more detailed study of these processes it is convenient to apply a
diagramatic approach to SU(3), describing amplitudes in terms of SU(3) flavor
flow topologies \cite{GHLR}. In this description, the two amplitudes of a
pair of  U-spin related processes have similar expressions, differing only
by their CKM factors. This proves that corresponding CP rate differences
have equal magnitudes and opposite signs \cite{He}.
Strangeness changing decay amplitudes are denoted by primes, while
strangeness conserving processes involve no primes. For instance,
\beqn\label{kpi}
A(B^0 \to K^+\pi^-) &=& -P' - T' - \frac{2}{3}P'^c_{EW}~,\cr
A(B_s \to \pi^+ K^-) &=& -P - T - \frac{2}{3}P^c_{EW}~.
\eeqn
Thus, we write only expressions for $\Delta S=1$ processes \cite{GHLR}:
\beqn\label{amps}
A(B^0 \to K^+\pi^-) &=& -P' - T' - \frac{2}{3}P'^c_{EW}~,\cr
A(B_s \to K^+ K^-) &=& -P' - T' -\frac{2}{3}P'^c_{EW} - PA' - E'~,\cr
\s A(B^0 \to K^0\pi^0) &=& P' - \frac{1}{3}P'^c_{EW} - P'_{EW} - C'~,\cr
A(B^+ \to K^0 \pi^+) &=& P' - \frac{1}{3}P'^c_{EW}  + A'~,\cr
A(B_s \to K^0 \Kbar^0) &=& P' - \frac{1}{3}P'^c_{EW} + PA' ~,\cr
A(B_s \to\pi^+\pi^-) &=& - PA' -E'~.
\eeqn

In the convention of Eqs.~(\ref{Heff})$-$(\ref{dbar}) the above amplitudes
are decomposed into two sets of terms
containing $V^*_{cb}V_{cs}$ and $V^*_{ub}V_{us}$.
(Correspondingly, amplitudes in the second Eq.~(\ref{kpi}) involve
$V^*_{cb}V_{cd}$ and $V^*_{ub}V_{ud}$). The first set consists
of a QCD penguin $P'$, an electroweak penguin $P'_{EW}$, a
color-suppressed electroweak penguin $P'^c_{EW}$
and a penguin annihilation term $PA'$, while the second set contains a
tree $T'$, color-suppressed $C'$, annihilation $A'$ and exchange $E'$
amplitude. Amplitudes in the first set carry each a weak phase
${\rm Arg}(V^*_{cb}V_{cs})=0$, while the other four terms have
a phase $\gamma$.

Let us discuss briefly the magnitudes of various terms.
The amplitudes obey the following hierarchy relations
\beq
\vert P'\vert \gg \vert T' \vert \sim \vert P'_{EW} \vert \gg \vert C'\vert
\sim\vert P'^c_{EW} \vert~,
\eeq
where a hierarchy factor of about 0.2 or 0.3 describes the ratio of
sequential amplitudes. This hierarchy was anticipated \cite{GHLR,FLE}
from the corresponding CKM factors, a color factor, QCD and electroweak
loop factors. It is supported both by relating $B\to K\pi$ and $B\to\pi\pi$
data \cite{CLEO} using flavor SU(3) \cite{DGR} and by recent QCD calculations
\cite{KLS,BBNS}. The other three amplitudes, $PA',~A'$ and $E'$, in which the
spectator
quark participates in the interaction, are usually assumed to be small
\cite{GHLR} (see also \cite{KLS,BBNS}), unless strongly amplified
by final state rescattering \cite{resc}.

The dominant term in the amplitudes (\ref{amps}) is $P'$ \cite{DGR}, occuring
in the first five processes. These decays are expected to have comparable
branching ratios of order $10^{-5}$, as measured for
$B^0\to K^+\pi^-,~B^0\to K^0\pi^0$ and $B^+\to K^0\pi^+$ \cite{CLEO}.
The decay $B_s \to \pi^+\pi^-$ and the U-spin related process
$B^0 \to K^+K^-$
involve only the much smaller combination $PA' + E'$, and are anticipated
to have much lower rates. Neglecting rescattering, one estimates for $B^0
\to K^+K^-$ a branching ratio of order $10^{-8}$ \cite{KLS}. The present
experimental upper limit
\cite{CLEO} is two orders of magnitude higher. Lowering this limit by one or
two orders of magnitude would settle the question of little rescattering.
A way of bounding the rescattering amplitude $A'$ in $B^+ \to K^0\pi^+$ (or,
similarly, of bounding $PA'$ in $B_s \to K^0\Kbar^0$) was discussed in
\cite{Falk}.

Assuming in the following that $PA' + E'$ can indeed be neglected, one has
\beqn\label{SU3}
A(B_s \to K^+ K^-) &\approx & A(B^0 \to K^+\pi^-)\cr
A(B_s \to \pi^+ K^-) &\approx & A(B^0 \to \pi^+\pi^-)~.
\eeqn
Once one obtains stringent limits on $PA' + E'$ through  $B^0
\to K^+K^-$ and $B_s \to \pi^+\pi^-$, these two relations provide tests for
U-spin symmetry acting on the spectator quarks \cite{HLW}. They imply, of
course, that the CP rate differences of all four processes are equal.
Equal CP rate differences in $B^0 \to\pi^+\pi^-$ and $B^0\to K^+\pi^-$
(assuming $PA'+E'=0$) were discussed in \cite{DH}.

Deviations from equalities in (\ref{SU3}) (once $PA' + E'$ is sufficiently
bounded) measure U-spin symmetry breaking. In the approximation of
factorized hadronic amplitudes \cite{KLS,BBNS}, U-spin breaking is introduced
through the ratio of corresponding form factors,
$f=F_{B_s K}(m^2_K)/F_{B\pi}(m^2_K)\approx
F_{B_s K}(m^2_{\pi})/F_{B\pi}(m^2_{\pi})$,
$$
A(B_s\to K^+ K^-) = f A(B^0\to K^+\pi^-)~,
$$
\beq\label{su3}
A(B_s\to K^-\pi^+) =  f A(B^0\to \pi^+\pi^-)~.
\eeq
{\it The rates of these four processes can be used not only to determine the
U-spin breaking factor $f$, but also to check the
factorization assumption by finding equal ratios of amplitudes in the two
cases.}

A similar U-spin equality in the absence of large rescattering effects
(i.e., neglecting $PA' - A'$) holds between the amplitudes of $B^+\to K^0
\pi^+$ and $B_s \to K^0\Kbar^0$, and between
the corresponding U-spin related processes, $B^+\to\Kbar^0 K^+$
and $B^0 \to \Kbar^0 K^0$. Again, the rates of these four processes can be used
to test U-spin symmetry and to measure U-spin symmetry breaking.

A method of determining the weak phase $\gamma$, using
$B^+ \to K^0\pi^+$ and the U-spin related processes $B^0\to K^+\pi^-$ and
$B_s \to \pi^+ K^-$, was discussed in \cite{GR}. Here we
reiterate the most important features of this suggestion,
which can be explained simply in terms of our
general U-spin considerations leading to Eqs.~(\ref{s})$-$(\ref{dbar}).
As noted, the rates of the four processes, in our case $B^0\to K^+\pi^-,~
B_s \to \pi^+ K^-$ and their charge-conjugates, depend on the magnitudes
of two hadronic amplitudes, $P'+(2/3)P'^c_{EW}$ and $T'$, and on their
relative strong phase and weak phase $\gamma$. (Note that $\vert
P+(2/3)P^c_{EW}\vert=\tan\theta_c\vert P'+(2/3)P'^c_{EW}\vert$ while
$\vert T\vert =\tan^{-1}\theta_c \vert T'\vert$). The four rates obey a
U-spin equality (\ref{asym}) between the two rate asymmetries in
$B^0\to K^{\pm}\pi^{\mp}$ and $B_s \to \pi^{\pm} K^{\mp}$. Thus, to solve
for $\gamma$ one needs one more input. This input is provided by the
charge-averaged rate of $B^{\pm}\to K^0\pi^{\pm}$ in which rescattering
is neglected, as can be further justified by improving bounds on
$B^0\to K^+ K^-$ and $B_s\to\pi^+\pi^-$.

This method can be generalized to other pairs of U-spin related decays.
For instance, one may use $B^0\to K^{*+}\pi^-$,~$B_s\to \rho^+ K^-$ and
their charge-conjugates, complemented by information on the charge-averaged
rate of  $B^{\pm} \to K^{*0}\pi^{\pm}$. As we have shown, U-spin breaking
effects can be measured by comparing rates and asymmetries of corresponding
processes. Including such effects in the analysis would result in a more
precise determination of $\gamma$.

Another suggested method \cite{FL}, based on comparing time-dependent
CP asymmetries in $B^0 \to \pi^+ \pi^-$ and in $B_s \to K^+ K^-$ (for which
one must tag the flavor of the neutral mesons at time of production), can
also be generalized to other U-spin related decays. This includes
comparing $B^0\to K_S\pi^0$ with $B_s \to K_S\pi^0$ and comparing
$B^0\to K_S K_S$ with $B_s\to K_S K_S$.  One measures for both channels
a $\cos\Delta mt$ and a $\sin\Delta mt$ term, or alternatively an oscillation
amplitude and an oscillation phase.
These four quantities determine four unknowns:  the ratio of
two hadronic amplitudes involving $V^*_{ub}V_{us}$ and $V^*_{cb}V_{cs}$,
a strong final-state phase between the two amplitudes, and two weak phases
$\beta$ and $\gamma$. U-spin breaking effects, represented in the
factorization approximation by certain ratios of form factors and decay
constants, cancel one another in the asymmetries.

To conclude, we pointed out several interesting properties of pairs of
U-spin related hadronic and radiative charmless $B^0,~B^+$ and $B_s$ decays.
We have
shown that in the U-spin symmetry limit every two such processes involve CP
rate differences which have equal magnitudes and opposite signs. Observing
large deviations from this prediction would be evidence for new physics.
We found six such pairs among decays to two light pseudoscalars, from which
information can be obtained about final state rescattering effects and
U-spin breaking. This information is required for an accurate determination of
the weak phase $\gamma$.

\vspace{0.3cm}
{\it Acknowledgments:\/}
I am grateful to the CERN Theory Division for its kind hospitality, and
would like to thank Gerhard Buchalla, Gil Paz, Dan Pirjol
and Jonathan Rosner for useful discussions. This work was supported in part by
the Israel Science Foundation founded by the Israel Academy of Sciences and
Humanities and by the United States - Israel Binational Science Foundation
under Research Grant Agrement 98-00237.

\newpage
\def \ajp#1#2#3{Am.~J.~Phys.~{\bf#1} (#3) #2}
\def \apny#1#2#3{Ann.~Phys.~(N.Y.) {\bf#1} (#3) #2}
\def \app#1#2#3{Acta Phys.~Polonica {\bf#1} (#3) #2 }
\def \arnps#1#2#3{Ann.~Rev.~Nucl.~Part.~Sci.~{\bf#1} (#3) #2}
\def \cmp#1#2#3{Commun.~Math.~Phys.~{\bf#1} (#3) #2}
\def \cmts#1#2#3{Comments on Nucl.~Part.~Phys.~{\bf#1} (#3) #2}
\def \cn{Collaboration}
\def \corn93{{\it Lepton and Photon Interactions:  XVI International Symposium,
Ithaca, NY August 1993}, AIP Conference Proceedings No.~302, ed.~by P. Drell
and D. Rubin (AIP, New York, 1994)}
\def \cp89{{\it CP Violation,} edited by C. Jarlskog (World Scientific,
Singapore, 1989)}
\def \dpff{{\it The Fermilab Meeting -- DPF 92} (7th Meeting of the American
Physical Society Division of Particles and Fields), 10--14 November 1992,
ed. by C. H. Albright \ite~(World Scientific, Singapore, 1993)}
\def \dpf94{DPF 94 Meeting, Albuquerque, NM, Aug.~2--6, 1994}
\def \efi{Enrico Fermi Institute Report No. EFI}
\def \el#1#2#3{Europhys.~Lett.~{\bf#1} (#3) #2}
\def \epjc#1#2#3{Eur.~Phys.~J.~C~{\bf#1} (#3) #2}
\def \f79{{\it Proceedings of the 1979 International Symposium on Lepton and
Photon Interactions at High Energies,} Fermilab, August 23-29, 1979, ed.~by
T. B. W. Kirk and H. D. I. Abarbanel (Fermi National Accelerator Laboratory,
Batavia, IL, 1979}
\def \hb87{{\it Proceeding of the 1987 International Symposium on Lepton and
Photon Interactions at High Energies,} Hamburg, 1987, ed.~by W. Bartel
and R. R\"uckl (Nucl. Phys. B, Proc. Suppl., vol. 3) (North-Holland,
Amsterdam, 1988)}
\def \ib{{\it ibid.}~}
\def \ibj#1#2#3{~{\bf#1} (#3) #2}
\def \ichep72{{\it Proceedings of the XVI International Conference on High
Energy Physics}, Chicago and Batavia, Illinois, Sept. 6--13, 1972,
edited by J. D. Jackson, A. Roberts, and R. Donaldson (Fermilab, Batavia,
IL, 1972)}
\def \ijmpa#1#2#3{Int.~J.~Mod.~Phys.~A {\bf#1} (#3) #2}
\def \ite{{\it et al.}}
\def \jhep#1#2#3{JHEP~{\bf#1} (#3) #2}
\def \jmp#1#2#3{J.~Math.~Phys.~{\bf#1} (#3) #2}
\def \jpg#1#2#3{J.~Phys.~G {\bf#1} (#3) #2}
\def \lkl87{{\it Selected Topics in Electroweak Interactions} (Proceedings of
the Second Lake Louise Institute on New Frontiers in Particle Physics, 15--21
February, 1987), edited by J. M. Cameron \ite~(World Scientific, Singapore,
1987)}
\def \ky85{{\it Proceedings of the International Symposium on Lepton and
Photon Interactions at High Energy,} Kyoto, Aug.~19-24, 1985, edited by M.
Konuma and K. Takahashi (Kyoto Univ., Kyoto, 1985)}
\def \mpla#1#2#3{Mod.~Phys.~Lett.~A {\bf#1} (#3) #2}
\def \nc#1#2#3{Nuovo Cim.~{\bf#1} (#3) #2}
\def \npb#1#2#3{Nucl.~Phys.~B {\bf#1} (#3) #2}
\def \pisma#1#2#3#4{Pis'ma Zh.~Eksp.~Teor.~Fiz.~{\bf#1} (#3) #2[JETP Lett.
{\bf#1} (#3) #4]}
\def \pl#1#2#3{Phys.~Lett.~{\bf#1} (#3) #2}
\def \plb#1#2#3{Phys.~Lett.~B {\bf#1} (#3) #2}
\def \pr#1#2#3{Phys.~Rev.~{\bf#1} (#3) #2}
\def \pra#1#2#3{Phys.~Rev.~A {\bf#1} (#3) #2}
\def \prd#1#2#3{Phys.~Rev.~D {\bf#1} (#3) #2}
\def \prl#1#2#3{Phys.~Rev.~Lett.~{\bf#1} (#3) #2}
\def \prp#1#2#3{Phys.~Rep.~{\bf#1} (#3) #2}
\def \ptp#1#2#3{Prog.~Theor.~Phys.~{\bf#1} (#3) #2}
\def \rmp#1#2#3{Rev.~Mod.~Phys.~{\bf#1} (#3) #2}
\def \rp#1{~~~~~\ldots\ldots{\rm rp~}{#1}~~~~~}
\def \si90{25th International Conference on High Energy Physics, Singapore,
Aug. 2-8, 1990}
\def \slc87{{\it Proceedings of the Salt Lake City Meeting} (Division of
Particles and Fields, American Physical Society, Salt Lake City, Utah, 1987),
ed.~by C. DeTar and J. S. Ball (World Scientific, Singapore, 1987)}
\def \slac89{{\it Proceedings of the XIVth International Symposium on
Lepton and Photon Interactions,} Stanford, California, 1989, edited by M.
Riordan (World Scientific, Singapore, 1990)}
\def \smass82{{\it Proceedings of the 1982 DPF Summer Study on Elementary
Particle Physics and Future Facilities}, Snowmass, Colorado, edited by R.
Donaldson, R. Gustafson, and F. Paige (World Scientific, Singapore, 1982)}
\def \smass90{{\it Research Directions for the Decade} (Proceedings of the
1990 Summer Study on High Energy Physics, June 25 -- July 13, Snowmass,
Colorado), edited by E. L. Berger (World Scientific, Singapore, 1992)}
\def \stone{{\it B Decays}, edited by S. Stone (World Scientific, Singapore,
1994)}
\def \tasi90{{\it Testing the Standard Model} (Proceedings of the 1990
Theoretical Advanced Study Institute in Elementary Particle Physics, Boulder,
Colorado, 3--27 June, 1990), edited by M. Cveti\v{c} and P. Langacker
(World Scientific, Singapore, 1991)}
\def \yaf#1#2#3#4{Yad.~Fiz.~{\bf#1} (#3) #2 [Sov.~J.~Nucl.~Phys.~{\bf #1} (#3)
#4]}
\def \zhetf#1#2#3#4#5#6{Zh.~Eksp.~Teor.~Fiz.~{\bf #1} (#3) #2 [Sov.~Phys. -
JETP {\bf #4} (#6) #5]}
\def \zpc#1#2#3{Zeit.~Phys.~C {\bf#1} (#3) #2}


\begin{thebibliography}{99}
\bibitem{U} U-spin is an SU(2) subgroup of flavor SU(3), under which the
$(d,s)$ pair of quarks is a doublet, similar to $(u,d)$ in isospin.
\bibitem{other} Our U-spin symmetry argument applies generally
to all cases in which one compares two processes where initial and
final states belong to common U-spin multiplets.
\bibitem{sign} Equal signs of CP asymmetries in two U-spin related processes
require that the sine of a relative strong phase changes sign under a U-spin
transformation.
\end{thebibliography}
\end{document}